\documentclass{article}
\usepackage{spconf,amsmath,graphicx}
\usepackage{multirow}
\usepackage{bm}
\usepackage{graphicx}
\usepackage{subfig}
\usepackage{enumitem,kantlipsum}

\title{Very Deep Convolutional Neural Networks for Raw Waveforms}
\name{Wei Dai*, Chia Dai*\thanks{*These authors contributed equally.}, Shuhui Qu, Juncheng Li, Samarjit Das}
\address{\{wdai,chiad\}@cs.cmu.edu, shuhuiq@stanford.edu, \{billy.li,samarjit.das\}@us.bosch.com}
\begin{document}
%
\maketitle
\begin{abstract}


Learning acoustic models directly from the raw waveform data with minimal processing is challenging. Current waveform-based models have generally used very few ($\sim$2) convolutional layers, which might be insufficient for building high-level discriminative features. In this work, we propose very deep convolutional neural networks (CNNs) that directly use time-domain waveforms as inputs. Our CNNs, with up to 34 weight layers, are efficient to optimize over very long sequences (e.g., vector of size 32000), necessary for processing acoustic waveforms. This is achieved through batch normalization, residual learning, and a careful design of down-sampling in the initial layers. Our networks are fully convolutional, without the use of fully connected layers and dropout, to maximize representation learning. We use a large receptive field in the first convolutional layer to mimic bandpass filters, but very small receptive fields subsequently to control the model capacity. We demonstrate the performance gains with the deeper models. Our evaluation shows that the CNN with 18 weight layers outperform the CNN with 3 weight layers by over 15\% in absolute accuracy for an environmental sound recognition task and matches the performance of models using log-mel features.

\end{abstract}
\vspace{-0.2cm}
\begin{keywords}
Convolutional Neural Networks, Raw Waveform, Acoustic Modeling, Neural Networks, Environmental Sound
\end{keywords}
\vspace{-0.6cm}
\section{Introduction}
\vspace{-0.2cm}
\label{sec:intro}

Acoustic modeling is traditionally divided into two parts: (1) designing a feature representation of the audio data, and (2) building a suitable predictive model based on the representation. However, it is often challenging and time-intensive to find the right representation in the so-called ``feature-engineering'' process, and the often heuristically designed features might not be optimal for the predictive task. Deep neural networks, which have achieved state-of-the-art performances in acoustic scene recognition~\cite{dcase_winner} and speech recognition~\cite{deep_speech}, have increasingly blurred the line between representation learning and predictive modeling. Instead of using the hand-tuned Gaussian Mixture Model features and Mel-frequency cepstrum coefficients, neural network models can directly take as input features such as spectrograms~\cite{deep_speech} and even raw waveforms~\cite{raw_speech}. By using simpler features, deep neural networks can be viewed as extracting feature representation {\it jointly} with classification, rather than separately~\cite{tara_raw}. This joint optimization is highly effective in speech recognition~\cite{deep_speech} and image classification~\cite{alexnet}, among others. 


A fundamental building block of these models is the convolutional neural networks (CNNs), which can learn spatially or temporally invariant features from pixels or time-domain waveforms. CNNs have famously achieved performance competitive or even surpassing human-level performance in the visual domains, such as object recognition~\cite{resnet} and face recognition~\cite{deepface, facenet}.
A common theme among these powerful CNN models is that they are usually very deep, with the number of layers ranging from tens to even over a hundred. Nonetheless, designing and training a deep network suitable for a new application domain remain challenging.


Recent works have applied CNNs to audio tasks such as environmental sound recognition and speech recognition and found that CNNs perform well with just the raw waveforms~\cite{raw_time_dnn, tara_raw, cnn_raw_time}. In one case, CNNs with time-domain waveforms can match the performance of models using conventional features like log-mel features~\cite{tara_raw}. These works, however, have mostly considered only less deep networks, such as two convolutional layers~\cite{tara_raw,piczak15}.


In this work, we propose and study very deep architectures with up to 34 weight layers, directly using time-series waveforms as the input. Our deep networks are efficient to optimize over long sequences (e.g., vector of length 32000), necessary for processing raw audio waveforms. Our architectures use a very small receptive field in the convolutional layers, but a large receptive field in the first layer chosen based on the audio sampling rate to mimic bandpass filter. Our models are fully convolutional, without fully connected layers and dropout, in order to maximize the representation learning in the convolutional layers and can be applied to audio of varying lengths. By applying batch normalization~\cite{batch_norm}, residual learning~\cite{resnet}, and a careful design of down-sampling layers, we overcome the difficulties in training very deep models while keeping the computation cost low.

On an environmental sound recognition task~\cite{urban_sound8k}, we show that deep networks improve the performance of networks with 2 convolutional layers by over 15\% in absolute accuracy. We further demonstrate that the performance of deep models using just the raw signal is competitive with models using log-mel features~\cite{piczak15}. To our knowledge, this is the first report of a parity performance between log-mel features and raw time signal for environmental sound recognition.


\vspace{-0.4cm}
\section{Very Deep Convolutional Networks}
\vspace{-0.2cm}
\begin{table*}[!ht]
\begin{minipage}[l]{0.64\textwidth}
\vspace{-0.3cm}
  \begin{tabular}{ |c|c|c|c|c| }
    \hline
    M3 (0.2M) & M5 (0.5M) & M11 (1.8M) & M18 (3.7M) & M34-res (4M) \\ \hline
    \multicolumn{5}{|c|}{Input: 32000x1 time-domain waveform} \\ \hline
    [80/4, 256] & [80/4, 128] & [80/4, 64] & [80/4, 64] & [80/4, 48] \\ \hline
     \multicolumn{5}{|c|}{Maxpool: 4x1 (output: 2000 $\times$ n)} \\ \hline
     [3, 256] & [3, 128] & [3, 64] $\times$ 2 & [3, 64] $\times$ 4 & \Big[\begin{tabular}{c}
  3, 48 \\
  3, 48 \\
  \end{tabular}\Big] $\times$ 3 \\ \hline
  \multicolumn{5}{|c|}{Maxpool: 4x1 (output: 500$\times$n)} \\ \hline
     &[3, 256] & [3, 128] $\times$ 2 & [3, 128] $\times$ 4 & \Big[\begin{tabular}{c}
  3, 96 \\
  3, 96 \\
  \end{tabular}\Big] $\times$ 4\\ \cline{2-5}
     & \multicolumn{4}{c|}{Maxpool: 4x1 (output: 125 $\times$ n)} \\ \cline{2-5}
     &[3, 512] & [3, 256] $\times$ 3 & [3, 256] $\times$ 4 & \Big[\begin{tabular}{c}
  3, 192 \\
  3, 192 \\
  \end{tabular}\Big] $\times$ 6\\ 
     \cline{2-5}
     & \multicolumn{4}{c|}{Maxpool: 4x1 (output: 32 $\times$ n)} \\ \cline{2-5}
     && [3, 512] $\times$ 2 & [3, 512] $\times$ 4 &\Big[\begin{tabular}{c}
  3, 384 \\
  3, 384 \\
  \end{tabular}\Big] $\times$ 3 \\ \hline
     \multicolumn{5}{|c|}{Global average pooling (output: 1 $\times$ n)} \\ \hline
     \multicolumn{5}{|c|}{Softmax} \\ \hline
  \end{tabular}
  \vspace{-0.3cm}
  \end{minipage}
  \vspace{-0.2cm}
  \begin{minipage}[l]{0.3\textwidth}
  \vspace{0.3cm}
  \caption{\small Architectures of proposed fully convolutional network for time-domain waveform inputs. M3 (0.2M) denotes 3 weight layers and 0.2M parameters. [80/4, 256] denotes a convolutional layer with receptive field 80 and 256 filters, with stride 4. Stride is omitted for stride 1 (e.g., [3, 256] has stride 1). [...] $\times k$ denotes $k$ stacked layers. Double layers in a bracket denotes a residual block and only occur in M34-res. Output size after each pooling is written as $m\times n$ where $m$ is the size in time-domain and $n$ is the number of feature maps and can vary across architectures. All convolutional layers are followed by batch normalization layers, which are omitted to avoid clutter. Without fully connected layers, we do not use dropout~\cite{dropout} in these architectures.}
  \label{tab:models}
  \end{minipage}
  \vspace{-0.3cm}
\end{table*}

Table~\ref{tab:models} outlines the 5 architectures we consider. Our architectures take as input time-series waveforms, represented as a long 1D vector, instead of hand-tuned features or specially designed spectrograms. Key design elements are:

\noindent{\bf Deep architectures.} To build very deep networks, we use very small receptive field 3 for all but the first 1D convolutional layers\footnote{Small receptive fields were first popularized by~\cite{vgg} for 2D images.}.
This reduces the number of parameters in each layer and control the model sizes and computation cost
as we go deeper. Furthermore, we aggressively reduce the temporal resolution in the first two layers by 16x with large convolutional and max pooling strides to limit the computation cost in the rest of the network~\cite{inception}. After the first two layers, the reduction of resolution is complemented by a doubling in the number of feature maps\footnote{In the visual domain this change in resolution and the number of features maps leads to more specialized filters at the higher layers (e.g., feature maps responding to faces) and more basic filters at the bottom (e.g., feature maps responding diagonal lines).}.  We use rectified linear units (ReLU) for lower computation cost, following~\cite{relu_speech, vgg}.

\noindent{\bf Fully convolutional networks.} Most deep convolutional networks for classification use 2 or more fully connected (FC) layers of high dimensions (e.g., 4096 in~\cite{vgg, alexnet}) for discriminative modeling, leading to a very high number of parameters. We hypothesize that most of the learning occurs in the convolutional layers, and with a sufficiently expressive representation from convolutional layers, no FC layer is necessary. We therefore adopt a {\it fully convolutional} design for our network construction~\cite{resnet, fcn}. Instead of FC layers, we use a single global average pooling layer which reduces each feature map into one float by averaging the activation across the temporal dimension. By removing FC layers, the network is forced to learn good representation in the convolutional layers, potentially leading to better generalization. We support this design decision in our evaluation and demonstrate that fully convolutional networks perform comparably or better compared with their counterparts endowed with FC layers.


\begin{figure}[!ht]
\vspace{-0.3cm}
\centering
\includegraphics[width=0.95\linewidth]{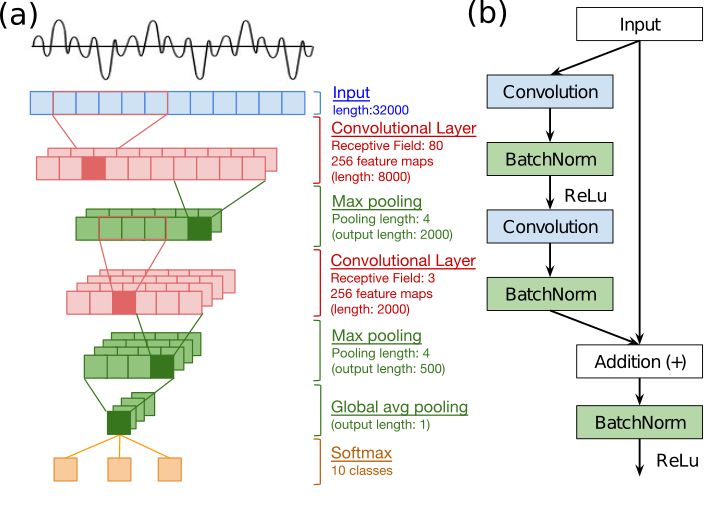}
\vspace{-0.4cm}
\caption{\small (a) The model architecture of M3 (Table~\ref{tab:models}). The input audio is represented by a single feature map (or channel). In each convolutional layer a feature map encodes activity level of the associated convolutional kernel. Note that the number of feature maps doubles as temporal resolution decreases by factor of 4 in the max pooling layers, capped by a global average pooling. Note that a reduction by factor of 4 in our max pooling layers is equal to a 2D max pooling with stride (2x2) used in many vision networks. (b) Residual block (res-block) used in M34-res. A resblock consists of two convolution layers.}
\vspace{-0.5cm}
\label{fig:panel}
\end{figure}

\noindent{\bf First layer receptive field.} Time-domain waveforms at a reasonable sampling rate (e.g. 8000Hz) over a few seconds could have very large number of samples along a single dimension. If we exclusively use small receptive field for all convolutional layers such as in~\cite{vgg}, which uses 3x3 in pixel for all layers, our model would need many layers in order to abstract high level features, which could be computationally expensive. Furthermore, audio sampling rate could affect the receptive field size in the first layer, since a field size of 80 at 8kHz sampling rate is at a different length scale than at 16kHz sampling rate. We thus choose our first layer receptive field to cover a 10-millisecond duration, which is similar to the window size for many MFCC computation. In Section~\ref{sec:exp} we show that a much smaller or larger receptive field gives poor performance.

\noindent{\bf Batch Normalization.}
We adopt auxiliary layers called batch normalization (BN)~\cite{batch_norm} that alleviates the problem of exploding and vanishing gradients, a common problem in optimizing deep architectures. BN normalizes the output of the previous layer so the gradients are well-behaved. This makes possible training very deep networks (M18, M34-res) that were not studied previously~\cite{very_deep_lvcsr}. Following~\cite{batch_norm}, we apply BN on the output of each convolutional layer before applying ReLU non-linearity.

\noindent{\bf Residual Learning.} Residual learning~\cite{resnet} is a recently proposed learning framework to ease the training of very deep networks. Normally we train a block of neural network layers to fit a desired mapping $\mathcal{H}(\bm{x})$ of $\bm{x}$ ($\bm{x}$ being the the input to the layers). In the residual framework, we instead let the block of layers approximate $\mathcal{F}(\bm{x})=\mathcal{H}(\bm{x})-\bm{x}$, the residual mapping. Residual learning is achieved through a skip connection in the residual block (``res-block'', Figure~\ref{fig:panel}b). We apply residual learning in M34-res (Table~\ref{tab:models}).


\vspace{-0.3cm}
\section{Experiment Details}
\label{sec:exp}
\vspace{-0.2cm}
We use UrbanSound8k dataset which contains 10 environmental sounds in urban areas, such as drilling, car horn, and children playing~\cite{urban_sound8k}. The dataset consists of 8732 audio clips of 4 seconds or less, totalling ~9.7 hours. We use the official fold 10 to be our test set, and the rest for training and validation. For computational speed, the audio waveforms are down-sampled to 8kHz and standardized to 0 mean and variance 1. We shuffle the training data but do not perform data augmentation. 

We train the CNN models using Adam~\cite{adam_opt}, a variant of stochastic gradient descent that adaptively tunes the step size for each dimension. We run each model for 100-400 epochs (defined as a pass over the training set) until convergence. The weights in each model are initialized from scratch without any pretrained model. We use glorot initialization~\cite{glorot_init} to avoid exploding or vanishing gradients. All weight parameters are subjected to $\ell_2$ regularization with coefficient 0.0001. Our models are implemented in Tensorflow~\cite{tensorflow} and trained on machines equipped with a Titan X GPU.

\noindent{\bf Additional Models.} To aid analysis, we train variants of models in Table~\ref{tab:models}.
The ``fc'' models replace global average pooling layer with 2 fully connected (FC) layers of dimension 1000 (Table~\ref{tab:fc}), since many conventional deep convolutional networks use 2 FC layers of dimension in the thousands ~\cite{alexnet, vgg, piczak15}. Following these works we also use a dropout layer between each fully connected layers for regularization, with a dropout rate of 0.3. We insert a batch normalization layer after each fully connected layers to aid training. These models have substantially more parameters than the original models due to the FC layers (Table~\ref{tab:fc}). Additionally, M3-big and M5-big (Table~\ref{tab:big}) are variants of M3 and M5, respectively, with 50\% and 100\% more filters (e.g., 384/256 filters in the first convolutional layer in M3-big/M5-big).

\vspace{-0.3cm}
\section{Results and Analyses}
\vspace{-0.2cm}
\begin{table}
\begin{center}
\begin{tabular}{ c | c | c  }
  \hline			
  Model & Test & Time\\ \hline
  M3 & 56.12\% & 77s \\ \hline
  M5 & 63.42\% & 63s \\ \hline
  M11 & 69.07\% & 71s \\ \hline
  M18 & 71.68\% & 98s \\ \hline
  M34-res & 63.47\% & 124s \\ 
  \hline  
\end{tabular}
\end{center}
\vspace{-0.4cm}
\caption{\small Test accuracies and training time per epoch (a sweep over the training set) for models in Table~\ref{tab:models} on UrbanSound8k dataset using a Titan X GPU.}
\vspace{-0.4cm}
\label{tab:results}
\end{table}

Table~\ref{tab:results} shows the test accuracies and training time for models in Table~\ref{tab:results}. We first note that M3 perform very poorly compared with the other models, indicating that 2-layered CNNs are insufficient to extract discriminative features from raw waveforms for sound recognition. This is in contrast with models using the spectrogram as input, which achieve good performance with just 2 convolutional layers~\cite{piczak15}, and shows that applying CNN directly on time-series data is challenging. M3-big, a variant of M3 with 50\% more filters and 2.5x more parameters, does not significantly improve the performance (Table~\ref{tab:big}), showing that shallow models have limited capacity to capture time-series inputs even with a larger model.

Deeper networks (M5, M11, M18, M34-res) substantially improve the performance. The test accuracy improves with increasing network depth for M5, M11, and M18. Our best model M18 reaches 71.68\% accuracy that is competitive with the reported test accuracy of CNNs on spectrogram input using the same dataset~\cite{piczak15}\footnote{Figure 4 in~\cite{piczak15} reports $\sim$68\% accuracy using a baseline CNN model. We point out that we have a different evaluation scheme: we use the 10-th fold as test set, while \cite{piczak15} performs 10-fold evaluation. Also we use sound at 8kHz sampling rate while they use the original 44.1kHz.}. The performance increases cannot be simply attributed to the larger number of parameters in the deep models. For example, M5-big has 2.2M parameters (Table~\ref{tab:big}) but only achieves 63.30\% accuracy, compared with the 69.07\% by M11 (1.8M parameters). By using a very deep architecture, M18 outperforms M3 by as much as 15.56\% in absolute accuracy, which shows that deeper architectures substantially improve acoustic modeling using waveforms. Furthermore, by using an aggressive down-sampling in the initial layers, very deep networks can be economical to train (Table~\ref{tab:results} Time column). When we use stride 1 instead of 4 in the first convolutional layer for M11, we observe a 3.5x increase in training time but a lower test accuracy (67.37\%) after 10 hours of training, compared with 68.42\% test accuracy reached in 2 hours by M18.

\begin{table}
\parbox{.40\linewidth}{
\centering
\begin{tabular}{ c | c   }
  \hline			
  Model & Test \\ \hline
  M11-srf & 64.78\% \\ \hline
  M18-srf & 65.55\% \\ \hline
  M11-lrf & 65.67\% \\ \hline
  M18-lrf & 65.08\% \\ \hline
\end{tabular}
\caption{\small Test accuracies for M11 and M18 variants different receptive field in the first convolutional layer. M11-srf and M18-srf have receptive field 8; M11-lrf and M18-lrf have 320.}
\vspace{-0.3cm}
\label{tab:srf}
}
\hfill
\parbox{.55\linewidth}{
\begin{tabular}{ c | c | c  }
  \hline			
  Model & Test & \# Parameters \\ \hline
  M3-big & 57.55\% & 0.5M \\ \hline
  M5-big & 63.30\% & 2.2M \\
  \hline  
\end{tabular}
\caption{\small Test accuracies for M3, M5 variants with more filters in the convolutional layers. M3-big, M5-big have 50\% and 100\% more filters (384 and 256 filters in the first layers, respectively).}
\vspace{-0.3cm}
\label{tab:big}
}
\end{table}


Interestingly, the performance improves with depth up to M18, at 71.68\% test accuracy. M34-res only achieves 63.47\% test accuracy. This is due to overfitting. We observe that with residual learning we have no problem optimizing deep networks like M34-res, and M34-res reaches an extremely high training accuracy of 99.21\%, compared with 96.72\% training accuracy by M18. We also observe overfitting in a residual variant of M11 network (not shown here) which reaches higher training accuracy but a lower test accuracy (by 0.17\%). Overfitting caused by very deep networks is well documented~\cite{resnet}. We believe that our dataset is too small to train M34-res without further regularization. Nonetheless, M34-res still outperforms M3 and M5.

We compare our fully convolutional network with conventional networks that use large fully connected layers (FC) for classification. Table~\ref{tab:fc} shows that FC layers can increase number of parameters significantly and increase training time by 2$\sim$95\%. However, FC layers do not improve test accuracy, and in the cases of M3-fc and M11-fc the additional FC layers lead to lower test accuracy (i.e., poorer generalization). We believe that the lack of FC layers in our network design pushes learning down to convolutional layers, leading to better representation and generalization.

To understand the effect of the receptive field (RF) size in the first convolutional layer, we train M11-srf and M18-srf, variants of M11 and M18 with RF 8, and M11-lrf and M18-lrf with RF 320. Table~\ref{tab:srf} shows that the performance degrades significantly by up to 6.6\% compared with M11 and M18 with RF 80. 
Previous works have shown that the first convolutional layer, when trained on raw waveforms, mimics wavelet transforms~\cite{raw_time_dnn, tara_raw}. Our results suggest that a small RF popularized by vision models is insufficient to capture the necessary bandpass filter characteristics in the first convolutional layer, while a large RF smooths out local structures and cannot effectively detect local impulse patterns.

We study the effect of batch normalization (BN) in optimizing very deep networks (Table~\ref{tab:no-bn}). Without BN, both M11-no-bn and M18-no-bn can be optimized to high training accuracy. Note that M18-no-bn results in lower test accuracy, indicating that BN has a regularization effect~\cite{batch_norm}. M34-no-bn could not be optimized without BN and performs close to random guess (10\%) after 159 epochs of training.

\begin{figure}[!ht]
\centering
\vspace{-0.3cm}
\includegraphics[width=0.48\textwidth]{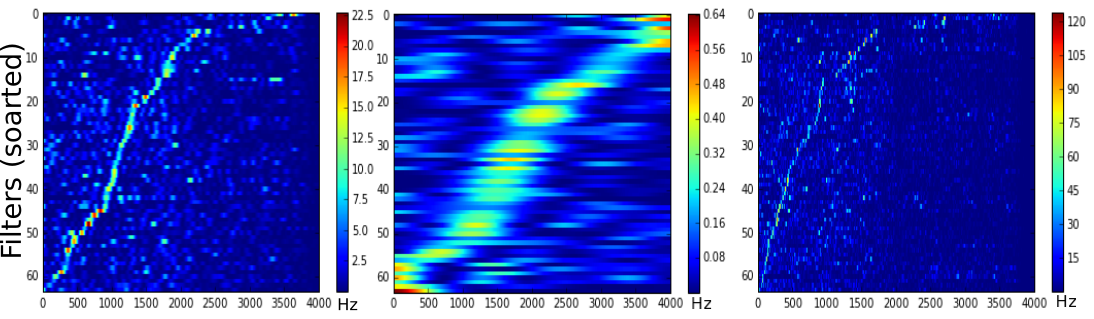}
\vspace{-0.5cm}
\caption{\small Kernels of the first convolutional layer after Fourier transformation, sorted by activation frequencies. Left: M18. Middle: M18-srf (small receptive field). Right: M18-lrf (large receptive field).}
\vspace{-0.5cm}
\label{fig:vis}
\end{figure}

Fig.~\ref{fig:vis} shows the learned kernels for M18 variants with different RF sizes in the first convolution layer. All of them learn a filter bank of bandpass filter. M18 (Fig.~\ref{fig:vis} left) has well-distributed filters. In contrast, the small RF model (Fig.~\ref{fig:vis} middle) has much more dispersed bands, and thus lower frequency resolution for subsequent layers. Conversely, large RF model (Fig.~\ref{fig:vis} right) has fine-grained filters, but does not have sufficient filters in the high frequency range, showing that it cannot effectively respond to local high frequency impulses.


\begin{table}
\vspace{-0.2cm}
\begin{center}
\begin{tabular}{ c | c | c | c }
  \hline			
  Model & Test & \# Parameters & Time \\ \hline
  M3-fc & 46.82\% & 129M & 150s\\ \hline
  M5-fc & 62.76\% & 18M & 66s \\  \hline  
  M11-fc & 68.29\% & 1.8M & 73s\\  \hline
  M18-fc & 64.93\% & 8.7M & 100s \\  \hline
\end{tabular}
\end{center}
\vspace{-0.4cm}
\caption{\small Test accuracy for models in Table~\ref{tab:models} endowed with fully connected (FC) layers. Time is training time per epoch.}
\label{tab:fc}
\end{table}

\begin{table}
\begin{center}
\begin{tabular}{ c | c | c  }
  \hline			
  Model & Train & Test \\ \hline
  M11-no-bn & 98.58\% & 69.38\% \\ \hline
  M18-no-bn & 99.33\% & 62.48\% \\
  \hline  
  M34-no-bn & 10.96\% & 11.45\% \\
  \hline 
\end{tabular}
\end{center}
\vspace{-0.4cm}
\caption{\small Test accuracies for models variants without batch normalization.}
\vspace{-0.5cm}
\label{tab:no-bn}
\end{table}

\vspace{-0.3cm}
\section{Conclusion}
\vspace{-0.2cm}
In this work, we propose very deep convolutional neural networks that operate directly on acoustic waveform inputs. Our networks, up to 34 weight layers, are efficient to optimize, thanks to the combination of batch normalization, residual learning, and down-sampling. We use a broad receptive field (RF) in the first convolutional layer and narrow RFs in the rest of the network. Our results show that a deep network with 18 weight layers outperforms networks with 2 convolutional layers by 15.56\% accuracy absolutely and achieves $71.8\%$ accuracy, competitive with CNNs using log-mel spectrogram inputs~\cite{piczak15}. Our fully convolutional networks compare favorably with those with fully connected layers.
Our proposed deep architectures hold the promise to improve CNNs for speech recognition and other time-series modeling.




\vfill\pagebreak

\vspace{-0.3cm}
\section{Acknowledgement}
\vspace{-0.2cm}
This work is supported by contract FA8702-15-D-0002 with Software Engineering Institute, a center sponsored by the United States Department of Defense.


\bibliographystyle{IEEEbib}
\bibliography{icassp}

\end{document}